# Generation and detection of pure valley current by electrically induced Berry curvature in bilayer graphene

Y. Shimazaki[1], M. Yamamoto[1,2], I. V. Borzenets[1], K. Watanabe[3], T. Taniguchi[3], and S. Tarucha[1,4]

[1]Department of Applied Physics, University of Tokyo, Bunkyo-ku, Tokyo 113-8656, Japan

[2]PRESTO, JST, Kawaguchi-shi, Saitama 331-0012, Japan

[3]National Institute for Materials Science, Tsukuba-shi, Ibaraki 305-0044, Japan

[4]Center for Emergent Matter Science (CEMS), RIKEN, Wako-shi, Saitama 351-0198, Japan

**Valley is a useful degree of freedom for non-dissipative electronics since valley current that can flow even in an insulating material does not accompany electronic current. We use dual-gated bilayer graphene in the Hall bar geometry to electrically control broken inversion symmetry or Berry curvature as well as the carrier density to generate and detect the pure valley current. We find a large nonlocal resistance and a cubic scaling between the nonlocal resistance and the local resistivity in the insulating regime at zero-magnetic field and 70 K as evidence of the pure valley current. The electrical control of the valley current in the limit of zero conductivity allows non-dissipative induction of valley current from electric field and thus provides a significant contribution to the advancement of non-dissipative electronics.**

Charge and spin are both well-defined quantum numbers in solids. Spintronics is a technology that uses the spin degree of freedom. The application range of spintronics has been largely expanded by the development of electrical techniques for generating and detecting the spin current (*1*). Valley is a quantum number defined in an electronic system whose band structure contains energetically degenerate but non-equivalent structures due to a certain crystal structure.

The valley degree of freedom can be handled by controlling occupation of the non-equivalent structures in the band providing a novel concept of so-called valleytronics. Valleytronics has recently been attracting growing interest as a promising the next generation of electronics, because non-dissipative pure valley current with accompanying no net charge flow can be manipulated for computational use. (As is the case for non-dissipative spin current in spintronics.) However, the field of valleytronics has remained largely unexplored. Electrical generation and detection of pure valley current is therefore the first step towards realization of valleytronics.

Among various material candidates for valleytronics, spatial inversion symmetry broken two-dimensional (2D) honeycomb lattice systems such as gapped graphene and transition metal dichalcogenide (TMDC) are predicted to be the most useful. These systems have two valleys called K and K′. Optical(*2–5*), magnetic(*6–9*), and electrical control of the valley(*10–12*) has been demonstrated. Particularly, Berry curvature which emerges in these spatial inversion symmetry broken honeycomb lattice systems enables electrical control of valley.

Berry curvature acts as an out of plane pseudo-magnetic field in momentum space and has opposite sign between the two valleys. Therefore, a transverse pure valley current is generated via anomalous velocity, in analogy to a transverse electronic current being generated via Lorentz force due to a magnetic field in real space (*13, 14*) (see Fig. 1D). This phenomenon is called valley Hall effect (*10*) and can be used to generate valley current. Inverse valley Hall effect, which converts the valley current into the transverse electric field, allows for detection of the pure valley current.

The valley Hall effect was first reported for photo-generated electrons in monolayer $MoS_2$ (*11*). However, small inter-valley scattering length in this material prevents detection of the pure valley current that does not accompany the electronic current. Compared to TMDC, graphene has much larger inter-valley scattering length owing to its higher crystal quality. Monolayer graphene on h-BN has more recently been used to generate and detect pure valley current, where crystal direction of the graphene was aligned to that of the h-BN such that the superlattice potential imposed by the h-BN structurally breaks the spatial inversion symmetry (*12*). The valley Hall effect was analyzed in detail with the carrier density as a parameter using metallic samples whose resistivity decreases with lowering the temperature, but leaving unaddressed the insulating regime which is more appropriate for investigating the pure valley current.

In this work we employed bilayer graphene (BLG) for generating and detecting the valley current. We used a perpendicular electric field to break the spatial inversion symmetry and to induce Berry curvature as well as a bandgap (see Fig. 1A). Dual-gated structure seen in Fig. 1B allows for electrical and independent control of the perpendicular electric field and the carrier density (*15–19*). This is in contrast with the monolayer graphene samples in (*12*), where the monolayer graphene has to be structurally aligned with h-BN via the processing of mechanical transfer. BLG valley Hall devices therefore have a higher potential in terms of tunability of the valley current and applications to electronic devices. Indeed we show that independent control of the Fermi level and the bandgap enables us to prove existence of the valley Hall effect in the insulating regime (where the local resistivity increases with lowering temperature). The significant advantage of the insulating system is that conversion from the electric field to the valley current is less dissipative than that in the metallic regime as a much smaller current is injected. Such a regime has not been accessible with conventional spin or valley Hall systems.

In bilayer graphene with broken spatial inversion symmetry, Berry curvature $\Omega$ and intrinsic valley Hall conductivity $\sigma_{xy}^{\text{VH}}$ are calculated as a function of Fermi energy $E_\text{F}$ in (*10, 20, 21*):

$$\Omega(E_\text{F}) = \tau_z \frac{\hbar^2}{m} \frac{\Delta\sqrt{E_\text{F}^2 - \Delta^2}}{E_\text{F}^3}, \qquad (1)$$

and

$$\sigma_{xy}^{VH}(E_F) = \begin{cases} \dfrac{4e^2}{h}\dfrac{\Delta}{|E_F|} & (|E_F| \geq \Delta) \\ \dfrac{4e^2}{h} & (|E_F| < \Delta) \end{cases}, \qquad (2)$$

where $\tau_z$ is the valley index ($\tau_z = -1$ for K and $+1$ for K′) and $m$ is the effective mass in BLG without spatial symmetry breaking. The Berry curvature $\Omega$ is only defined for $|E_F| \geq \Delta$ (half the bandgap, see Fig. 1A). $\sigma_{xy}^{VH}$ saturates at the maximum value $4e^2/h$ when the Fermi level lies in the gap, because all occupied states in the valence band contribute to the valley Hall effect. Away from the gap, for example, when the Fermi energy lies above the gap, the conduction band (which has opposite sign of Berry curvature with the valence band), contributes to reduce $\sigma_{xy}^{VH}$.

To detect the pure valley current, nonlocal resistance $R_{NL}$ was measured in the same scheme as is widely used in the spintronics field to detect pure spin current (*22–26*). We observed three orders of larger $R_{NL}$ at the charge neutrality point in the presence of a perpendicular electric field than $R_{NL}$ due to Ohmic contribution (explained later). We also found a cubic scaling relation between $R_{NL}$ and resistivity $\rho\,(=1/\sigma_{xx})$ which is expected to appear when $\sigma_{xx}$ is much larger than $\sigma_{xy}^{VH}$ in the intrinsic valley Hall effect. This cubic scaling was reproduced in multiple devices. From these findings we conclude that the origin of the observed large nonlocal resistance is the transport mediated by pure bulk valley current in a gapped state with electrically induced Berry curvature.

Fig. 1B and C shows the schematic of the dual-gated BLG device and the AFM image of the device, respectively. BLG is encapsulated between two h-BN layers (*27*) [(*28*), section 1.1] and gated through the h-BN layer from the top and from the bottom. The local and nonlocal resistance $R_L$ and $R_{NL}$ were derived from measurement of 4-terminal resistance $R_{ij,kl}$ which is defined by the voltage between terminals $i$ and $j$ divided by the charge current injected between terminals $k$ and $l$ (see Figs. 1 C and D). Unless mentioned, $R_L$ and $R_{NL}$ denotes $R_{57,38}$ and $R_{45,67}$, respectively. The measurement was performed at 70K using a low frequency (around 1Hz) lock-in technique, unless mentioned (*29*).

Figure 2A and B shows the gate voltage dependence of $R_L$ and $R_{NL}$, respectively. At the charge neutrality point (CNP), $R_L$ increases with the displacement field ($D$) (see Fig.2A), reflecting the bandgap opening due to inversion symmetry breaking(*15–19, 30*). We found that $R_{NL}$ also increases with $D$ around the CNP.

In analogy with the spin Hall effect (*23*) [(*28*), section 2.8], $R_{NL}$ arising from the valley Hall and inverse valley Hall effects is given by

$$R_{NL} = \frac{1}{2}\left(\frac{\sigma_{xy}^{VH}}{\sigma_{xx}}\right)^2 \frac{W}{\sigma_{xx} l_v} \exp\left(-\frac{L}{l_v}\right), \qquad (3)$$

where $\sigma_{xy}^{VH}$ and $l_v$ are the valley Hall conductivity and the inter-valley scattering length, respectively. $W$ and $L$ are the width and length of the Hall bar channel (see Fig. 1C). Local conductivity $\sigma_{xx}$ is minimized at CNP and with increasing $D$, thus enhancing $R_{NL}$ (Eq. 3). For a given $D$, $R_{NL}$ is additionally maximized around CNP due to the maximal valley Hall conductivity $\sigma_{xy}^{VH}$ (Eq. 2). We confirmed that $R_{NL}$ is unchanged when swapping the measurement terminals, i.e. $R_{45,67} \sim R_{67,45}$ [(*28*), section 2.2]. We also consider a contribution of trivial Ohmic resistance which is due to classical diffusive charge transport to the measured nonlocal resistance. The Ohmic contribution can be calculated using the van der Pauw formula $R_{NL} = \frac{\rho}{\pi}\exp\left(-\pi\frac{L}{W}\right)$ (*12, 24–26*), where we define resistivity $\rho = R_L(W/L)$, and is compared with the measured nonlocal resistance in Fig. 2C. The measured $R_{NL}$ is three orders of magnitude larger than the calculated Ohmic contribution. We therefore exclude the Ohmic contribution as the origin of the observed $R_{NL}$.

In the gapped BLG, the electron conduction mechanism depends on the temperature $T$. At high $T$ it is dominated by thermal activation across the bandgap, namely band transport, while at low $T$ by hopping conduction between impurity states (*15, 17–19*).

The temperature dependence of maximum $\rho$ with respect to carrier density ($\rho^{max}$) was measured for various displacement fields $D$ (Figure 3A, inset). We plot $1/\rho^{max}$ as a function of $1/T$ for $D = 0.55\,\text{V/nm}$ as a typical example in Fig. 3A. The temperature dependence is strong at high $T$ (>79K) reflecting band conduction and weak at low $T$ reflecting hopping conduction. The temperature dependence over the whole range is reproduced well by a double exponential function:

$$\frac{1}{\rho^{max}} = \frac{1}{\rho_1}\exp\left(-\frac{E_1^L}{k_B T}\right) + \frac{1}{\rho_2}\exp\left(-\frac{E_2^L}{k_B T}\right), \qquad (4)$$

Where $E_1^L$ ($E_2^L$) and $\rho_1$ ($\rho_2$) are the activation energy and the local resistivity, respectively for the high (low) $T$ regime. $2E_1^L$ indicates the bandgap size, and is around 80meV at the highest $D$ [(28), section 2.4]. The crossover temperature $T_c$ between the high and low temperature regions is determined by the crossing point of the first and second term of Eq. 4 as shown in Fig. 3A. The temperature dependence of the maximum nonlocal resistance $R_{NL}^{max}$ was also measured (Fig. 3B, inset) and analyzed with the following fitting function in the same way as for $1/\rho^{max}$ $\rho^{max}$ as shown in Fig. 3B:

$$\frac{1}{R_{NL}^{max}} = \frac{1}{R_1}\exp\left(-\frac{E_1^{NL}}{k_B T}\right) + \frac{1}{R_2}\exp\left(-\frac{E_2^{NL}}{k_B T}\right), \qquad (5)$$

where $E_1^{NL}$ ($E_2^{NL}$) are the activation energy and $R_1$ ($R_2$) are fitted proportionality factors, respectively for the high (low) $T$ regime. The temperature dependence is quite similar to that of $\rho^{max}$ in Fig. 3A. We also plotted the crossover temperature $T_c$ for both $1/\rho^{max}$ and $1/R_{NL}^{max}$ as a function of $D$ in Fig. 3C. The $T_c$ for divides the displacement field-temperature plane into the band conduction region (light green) and the hopping conduction region (light red).

The critical temperatures $T_c$ of $1/\rho^{max}$ and $1/R_{NL}^{max}$ coincide for $D > 0.4$V/nm, indicating that there is correlation of the crossover behavior between the local and nonlocal transport. However, it deviates for $D < 0.4$V/nm or the following two possible reasons. The first possible reason is underestimation of the $T_c$ of the nonlocal transport in the low $D$ region as the nonlocal voltage becomes significantly small at high $T$ and low $D$, making precise measurement of $R_{NL}$ difficult. The second possible reason is that the nonlocal transport via valley current is less affected by charge puddles compared to the local transport although we do not yet fully understand the reason for it.

One noticeable result is that the $T_c$ of the nonlocal transport almost linearly depends on $D$ for the entire region in Fig. 3D (see the blue curve) [(*28*), section 2.5]. This behavior may indicate that the $T_c$ is affected by the size of bandgap but less affected by the size of potential fluctuation by charge puddles. Note that all four of the fitting parameters in Eq.4,5 have $D$ dependence, therefore, obtaining an analytical relationship between $D$ and $T_c$ is not straightforward.

Another noticeable result is that the high-$T$ activation energy $E_1$ is different between the local and non-local transport [(*28*), section 2.4 and 2.5, see Fig. S7 and S8]. This already implies there is no linear relation between $R_{NL}$ and $\rho$ in our device. This observation is in contrast with the previous report on monolayer graphene (*12*), where both activation energies were similar.

We now present the scaling relation between $\rho$ and $R_{NL}$ at CNP. Fig. 4 is the plot of $R_{NL}$ vs. $\rho$ obtained from $R_{NL}$ and $\rho$ for various displacement fields $D$. The crossover behavior between the band conduction and the hopping conduction shows up again on this plot. In the band conduction region, we observe a clear cubic scaling relation (green line) while we observe saturation in the hopping conduction region. Note similar qubic and saturating scaling relations are obtained for different carrier densities [(*28*), section 2.6].

By assuming a constant inter-valley scattering length and replacing $\sigma_{xx}$ with $\rho^{-1}$ in Eq. 3, we derive the following scaling relation between $R_{NL}$ and $\rho$:

$$R_{NL} \propto (\sigma_{xy}^{VH})^2 \rho^3, \qquad (6)$$

The cubic scaling between $R_{NL}$ and $\rho$ holds for the constant valley Hall conductivity, which is expected when the Fermi level is in the bandgap or near CNP (see Fig. 1A) for the intrinsic valley Hall effect, $\sigma_{xy}^{VH} = 4e^2/h$. The observed cubic relation for the small $D$ in Fig. 4 is therefore consistent with the theoretical expectation, providing unambiguous evidence of the valley transport.

Using $\sigma_{xy}^{VH} = 4e^2/h$; plugging in $\sigma_{xx} = \rho^{-1}$ and the sample dimensions into Eq. 3 we obtain $l_v = 1.6 \mu m$. This is comparable to the estimated inter-valley scattering length in previous works (*12, 31*). By taking different sets of four terminals we observed a significantly increasing decay of $R_{NL}$ with $L$ [(*28*), section 2.3] probably due to the valley relaxation by the edge scattering as discussed in a weak localization study (*31*).

We here note that Eqs. 3 and 6 are only valid for $\sigma_{xx} \gg \sigma_{xy}^{VH}$ [(*28*), section 2.8]. Otherwise we need to solve the conductance matrix and the diffusion equation of the entire Hall bar in a self-consistent way. Indeed, deviation from the cubic scaling in the large $D$ region observed in Fig. 4 may come from inapplicability of Eqs. 3 and 6. But it does not account for the saturation of $R_{NL}$ for large $\rho$ [(*28*), section 2.8]. Another possible scenario to account for the saturating of $R_{NL}$ is the crossover of conduction mechanism as discussed in Fig. 3D. In anomalous Hall effect studies, the crossover between the metallic and the hopping transport regime has been experimentally studied, and the scaling relation $\sigma_{xy} \propto \sigma_{xx}^{1.6}$ has been reported in a wide range of materials (*13*). If we apply this experimental rule for Eq. 6, we find $R_{NL}$ to be almost constant with $\rho$. However, we are again cautious about the validity of Eq. 6 in this argument, because in the saturation region $\sigma_{xx} < \sigma_{xy}^{\text{VH}}$ for $\sigma_{xy}^{\text{VH}} = 4e^2/h$. However, by including extrinsic contributions, for example the side-jump contribution discussed in (*10*), $\sigma_{xy}^{\text{VH}}$ can be smaller than $4e^2/h$ and $\sigma_{xx}$. In such a case, we can keep the above-described analogy with the anomalous Hall effect. Further experimental and theoretical investigations are needed for the valley Hall effect in the insulating regime, where conventional formulas are not applicable.

We finally exclude another scenario to account for the $R_{\text{NL}}$ observed here. In the gap of bilayer graphene, presence of the localized states along the edge resulting from the topological property of BLG was predicted theoretically (*32*). This might also contribute to the nonlocal transport. In the large displacement field or large bandgap, the bulk shunting effect is small and the conduction becomes dominated by the edge transport. In such a case, $R_{\text{NL}}$ should be proportional to the local resistance obtained by a 4-terminal measurement [(*28*), section 2.7]. This linear scaling does not fit any of the observed features (demonstrated in Fig. 4).We draw a linear scaling line in blue in Fig. 4, but this does not fit any of the observed features. Even when we consider the effect of bulk shunting, we find that the scaling is far from the cubic line [(*28*), section 2.7]. So we exclude the possibility of edge transport as the origin of the observed $R_{\text{NL}}$ and conclude that it comes from the bulk valley current in the gap. (Also the transport through the localized states along the edge was denied by the measurement on Corbino geometry device) (*19*).

We used a dual-gated BLG in the Hall bar geometry to electrically control broken inversion symmetry of BLG and therefore valley degree of freedom. We observed a large nonlocal resistance in the insulating regime at 70 K and revealed a cubic scaling between the nonlocal resistance and local conductivity as an indication of pure valley current flow. The valley current is fully controlled by electrical gating (with the bandgap, the Fermi level and broken inversion symmetry as parameters). This will allow for further study on underlying physics of the valley current in particular for $\sigma_{xx} < \sigma_{xy}^{VH}$ as well as applications for non-dissipative electronics devices.

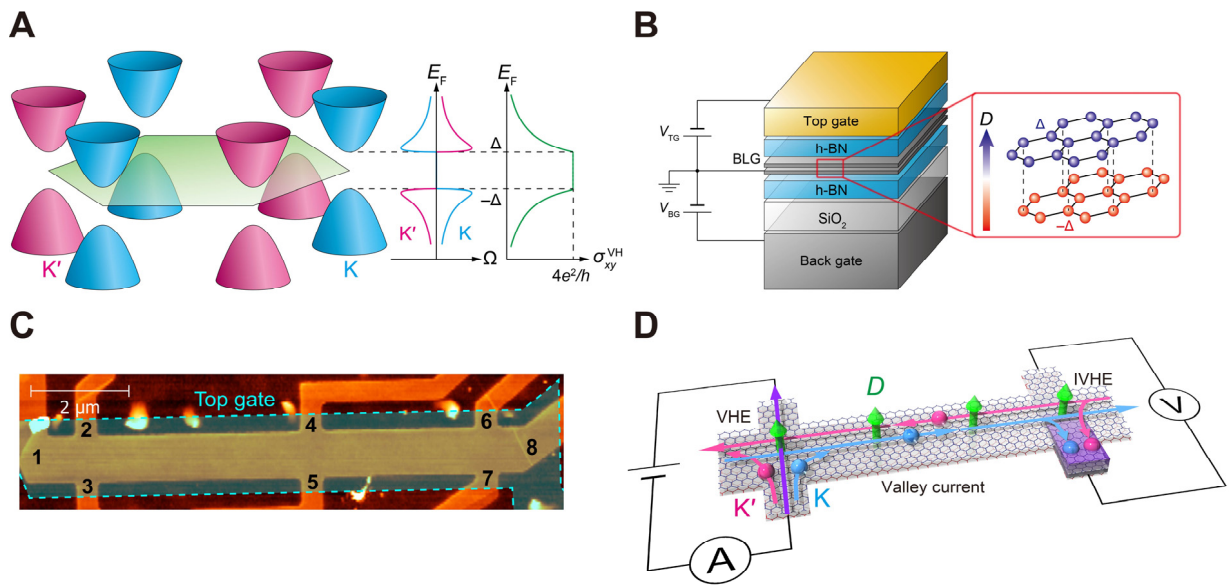

**Fig. 1. Detection scheme of nonlocal resistance due to valley current flow in BLG.**

(**A**) Band structure, Berry curvature, and valley Hall conductivity of BLG. Band gap $2\Delta$ and Berry curvature $\Omega$ emerge due to broken spatial symmetry. Valley Hall conductivity $\sigma_{xy}^{\text{VH}}$ which is calculated by integrating Berry curvature is constant in the bandgap.

(**B**) Schematic of the dual-gated BLG device. The top gate is a gold film, and the back gate is a p-doped silicon substrate. Two h-BN layers and a SiO2 film are used as gate insulators. By using these gates, the carrier density and perpendicular electric field (displacement field) are independently varied. (Balloon) Lattice structure of AB stacked BLG. In the presence of displacement field, energy difference between top and bottom layer emerges. Therefore spatial inversion symmetry is broken in this system, and Berry curvature as well as bandgap emerges.

(**C**) AFM image of the BLG devices without the top h-BN. The light blue box indicates the top gate area. The BLG has a mobility of ~15,000cm$^2$/Vs at both 1.5K and 70K.

(**D**) Schematic of the nonlocal resistance measurement and the nonlocal transport mediated by pure valley current. Electric field driving the charge current in the left generates a pure valley current in the transverse direction via valley Hall effect (VHE). This valley current is converted into electric field or nonlocal voltage in the right via inverse valley Hall effect (IVHE) to generate the nonlocal resistance.

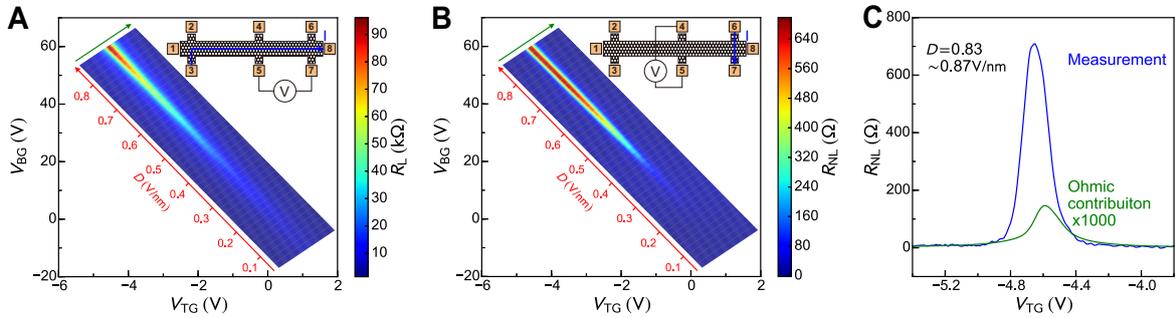

**Fig. 2. Measured local and nonlocal resistances $R_L$ and $R_{NL}$.**

(**A,B**) Gate voltage dependence of $R_L$ (**A**) and $R_{NL}$ (**B**). The displacement field from the back (top) gate is defined by $D_{BG} = \varepsilon_{BG}(V_{BG} - V_{BG}^0)/d_{BG}$, $(D_{TG} = -\varepsilon_{TG}(V_{TG} - V_{TG}^0)/d_{TG})$ where $\varepsilon_{BG}(\varepsilon_{TG})$ and $d_{BG}(d_{TG})$ are relative dielectric constant, and thickness of back (top) gate, respectively and $V_{BG}^0(V_{TG}^0)$ is the offset of the back (top) gate voltage under the top gated region due to environmental doping. The displacement field $D$ is defined by the average of $D_{BG}$ and

$D_{TG}$. The red axis shows the scale of $D$. (Inset) Schematic of the measurement configuration. The blue arrow shows the charge flow.

(**C**) Comparison of the measured $R_{NL}$ in blue with calculation of the Ohmic contribution (magnified 1000 times) in green. The $R_{NL}$ curve is extracted from the data along the green arrow in (**B**) at the highest $D$. The Ohmic contribution curve is calculated using the $R_L$ data along the green arrow in (**A**) at the highest $D$ (see text).

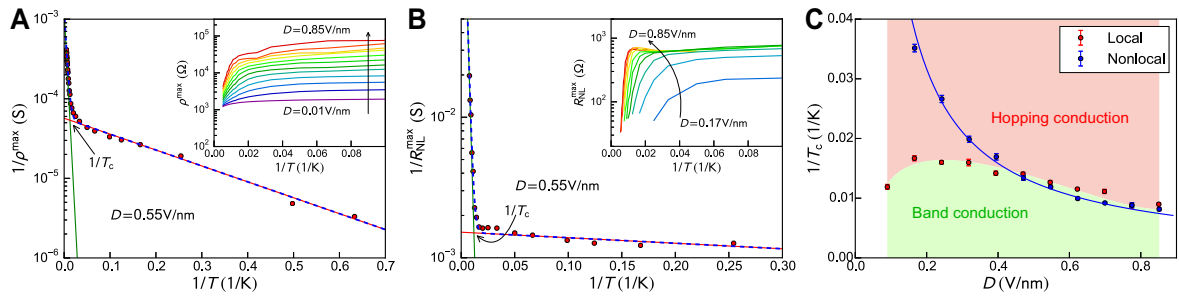

**Fig. 3. Temperature dependence of $\rho^{max}$ and $R_{NL}^{max}$.**

(**A**, **B**) Typical data fitting using a double exponential function for $1/\rho^{max}$ (**A**) and $1/R_{NL}^{max}$ (**B**). The blue broken curve indicates the fitting curve. The green (red) line indicates the contribution from band (hopping) conduction. The crossover temperature $T_c$ is defined by the crossing of the two lines. (Inset)Temperature dependence of maximum $\rho$ (**A**) and $R_{NL}$ (**B**) with respect to the carrier density. $D$ is varied from 0.85V/nm (red) to 0.01V/nm (purple) in (**A**) and from 0.85V/nm (red) to 0.17V/nm (blue) in (**B**) with a constant interval.

(**C**) $T_c$ derived from the data fitting as in (**A**) for $1/\rho^{max}$ and (**B**) for $1/R_{NL}^{max}$. The error bars come from the accuracy of the fitting. Light green (red) area is the region of band conduction (hopping conduction). The blue curve shows the fitting result for the nonlocal $1/T_c$ vs. $1/D$.

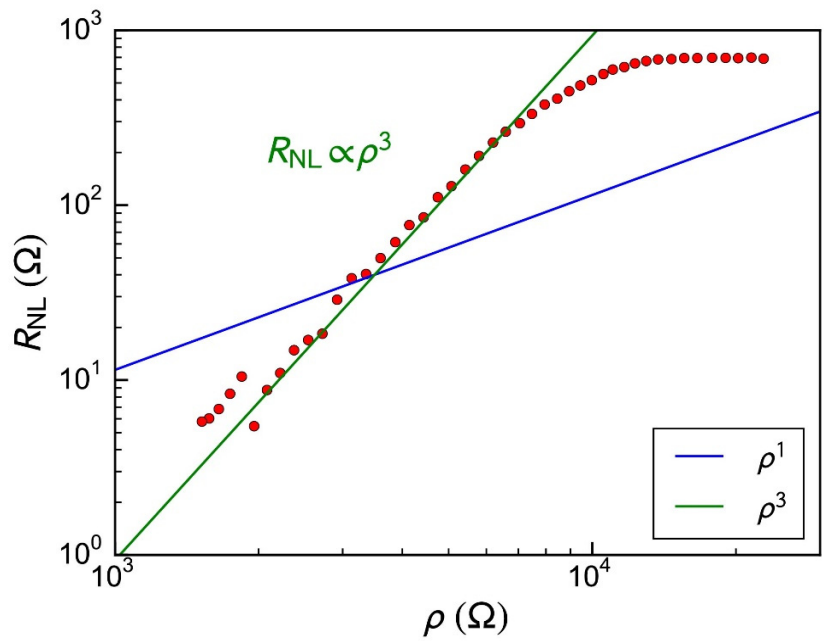

**Fig. 4. Scaling relation between $\rho$ and $R_{NL}$ at CNP.**

Each data point is extracted from Fig. 2A and B for a different $D$ value ranging from 0.01 to 0.85V/nm at CNP.

# Supplementary Material for

# Generation and detection of pure valley current by electrically induced Berry curvature in bilayer graphene

Y. Shimazaki, M. Yamamoto, I. V. Borzenets, K. Watanabe, T. Taniguchi, and S. Tarucha

Contents

1. **Materials and Methods**

**1.1 Device fabrication**

We used a mechanical exfoliation technique to prepare bilayer graphene (BLG) and h-BN flakes. The number of layers of each graphene flake on $SiO_2$/Si substrate was identified by optical contrast. The $SiO_2$ is 285nm thick, and the Si is heavily p-doped and used for back gating. We transferred the BLG flakes onto the h-BN flakes prepared on $SiO_2$ using a PMMA transfer technique reported in (*33*). Then Ti/Au (10nm/190nm) was deposited to make Ohmic contacts. BLG was etched into a Hall bar by Ar plasma. After each transfer and lithography step except for the step between the Ohmic contact deposition and Ar plasma etching, the device was annealed at 300°C in $Ar/H_2$ atmosphere for a few hours to remove the resist residue. However, the PMMA residue could not be completely removed by the anneal, so we used a mechanical cleaning technique (*34–36*) utilizing an AFM in tapping mode. After shaping the Hall bar, an h-BN flake was transferred on top of the BLG/h-BN stacking layer. Ti/Au (10nm/190nm) was finally deposited onto the h-BN/BLG/h-BN stacking layer to make the top gates. The thickness of the top and bottom h-BN layers measured by AFM was 21nm and 35nm, respectively.

## 2. Supplementary Text

### 2.1 Measurement error due to the current leakage

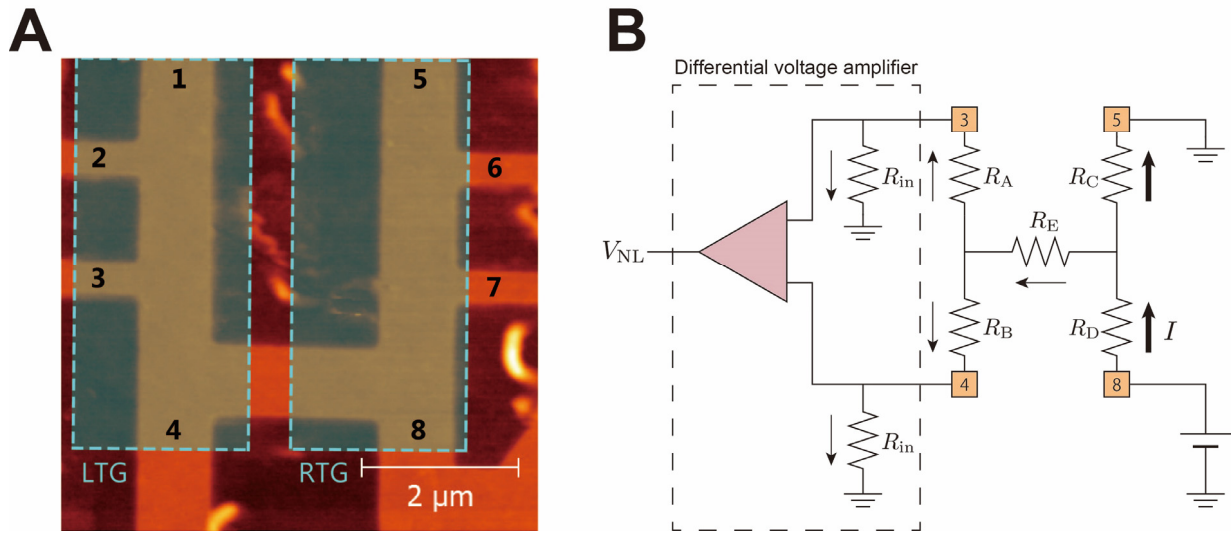

**Fig. S1: Asymmetric device and the resistor model which explains the measurement error due to current leakage.**

(**A**) The AFM image of the asymmetric device before encapsulation with an h-BN flake. The device has a dual gate structure with two h-BN insulating layers. The left and right region in light blue indicate the left top gate (LTG) and the right top gate (RTG), respectively both of which are placed after encapsulation with the h-BN flake.
(**B**) The circuit model of the measurement setup. This model describes the measurement error associated with the current leakage effect via the finite input impedance of the voltage preamplifier. The BLG device is represented as the five resistors ($R_A$ to $R_E$) connected with each other. Current $I$. is injected via $R_D$ into the BLG. A differential voltage amplifier, which has the input impedance $R_{in}$ between each input and the ground, is used to amplify the nonlocal voltage. Each arrow in the figure indicates the current flow.

A finite input impedance of a voltage amplifier which was used in the measurement circuit causes artifacts due to the current leakage which we should distinguish from the real signal. In Fig. S1A we show a device in which such a current leakage effect was clearly observed. The nonlocal resistance $R_{NL}$ was measured by injecting current from terminal 8 to 5 and measuring voltage difference between terminal 4 and 3. Local resistivity $\rho$ was measured simultaneously by detecting the voltage drop between terminal 7 and 6. The nonlocal voltage was measured after being amplified with a commercial differential voltage amplifier (SR560) with a 100 MΩ input impedance. Fig. S1B describes a simple circuit model for the device and measurement setup. There is a finite current flowing into the voltage amplifier because of the finite input impedance between the inputs and the ground. If the resistances $R_A$ and $R_B$ are not equal, a finite voltage difference between the two inputs emerges and is detected as a nonlocal voltage. At the charge neutrality point, the dominant source of $R_A$ and $R_B$ is the device resistance under the top gate, because the other device area is highly doped by the back gate voltage. Because of the asymmetry in the device geometry (Fig. S1A), $R_A$ and $R_B$ are not equivalent and therefore a finite nonlocal voltage reflecting the nonlocal resistance of

$$R_{NL} = \left( \frac{R_{in}}{R_{in} + R_B} - \frac{R_{in}}{R_{in} + R_A} \right) R_C \sim \frac{R_A - R_B}{R_{in}} R_C, \quad (S1)$$

emerges. There is a characteristic scaling relation between $R_{NL}$ and $\rho$, i.e. $R_{NL} \propto \rho^2$ in this situation because $R_A$, $R_B$ and, $R_C$ are almost proportional to the local resistivity $\rho$ of the device.

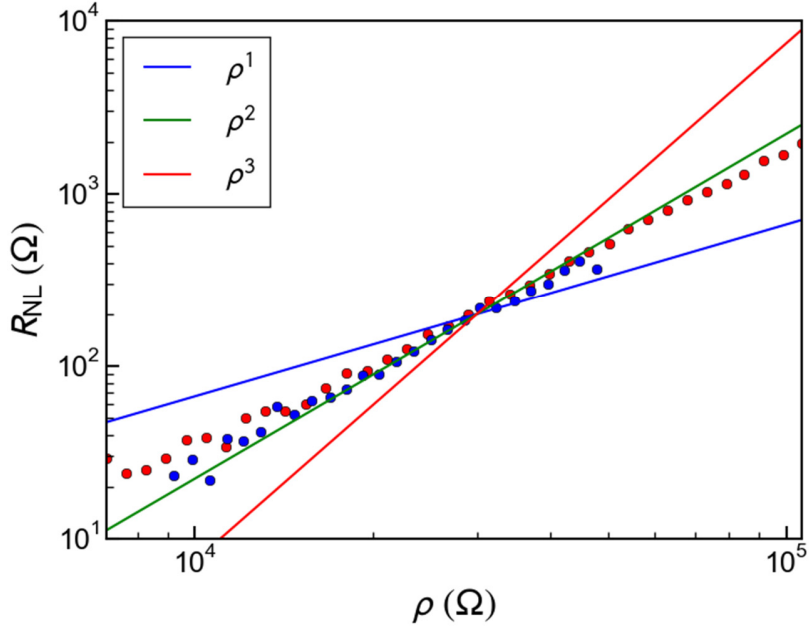

**Fig. S2: Scaling relation between $R_{NL}$ and $\rho$ in the asymmetric device.**

A scaling relation of $R_{NL}$ vs. $\rho$ as obtained at the charge neutrality point at 100K for various displacement fields in the same way as Fig. 4 in the main text. LTG and RTG voltages (see Fig. S1A) were equivalent and tuned simultaneously. $\rho$ was determined by the 4-terminal resistance $R_{76,85}$. The red (blue) points were obtained for the positive (negative) displacement field.

Fig. S2 shows the experimental observation of the square scaling relation between $R_{NL}$ and $\rho$ in the device shown in Fig. S1A. The measurement was performed with a DC setup. When $\rho \sim 100 k\Omega$. i.e. $R_C \sim 400 k\Omega$ and $R_A - R_B \sim$ a few $100 k\Omega$ from the aspect ratio, $R_{NL} \sim 10^3 \Omega$ is obtained from Eq. S1, which is consistent with the experimental result. On the other hand, neither cubic scaling nor saturation feature of $R_{NL}$ was observed. This is probably due to the reflection of valley current at the interface between the top-ungated doped region and the top-gated charge neutral region.

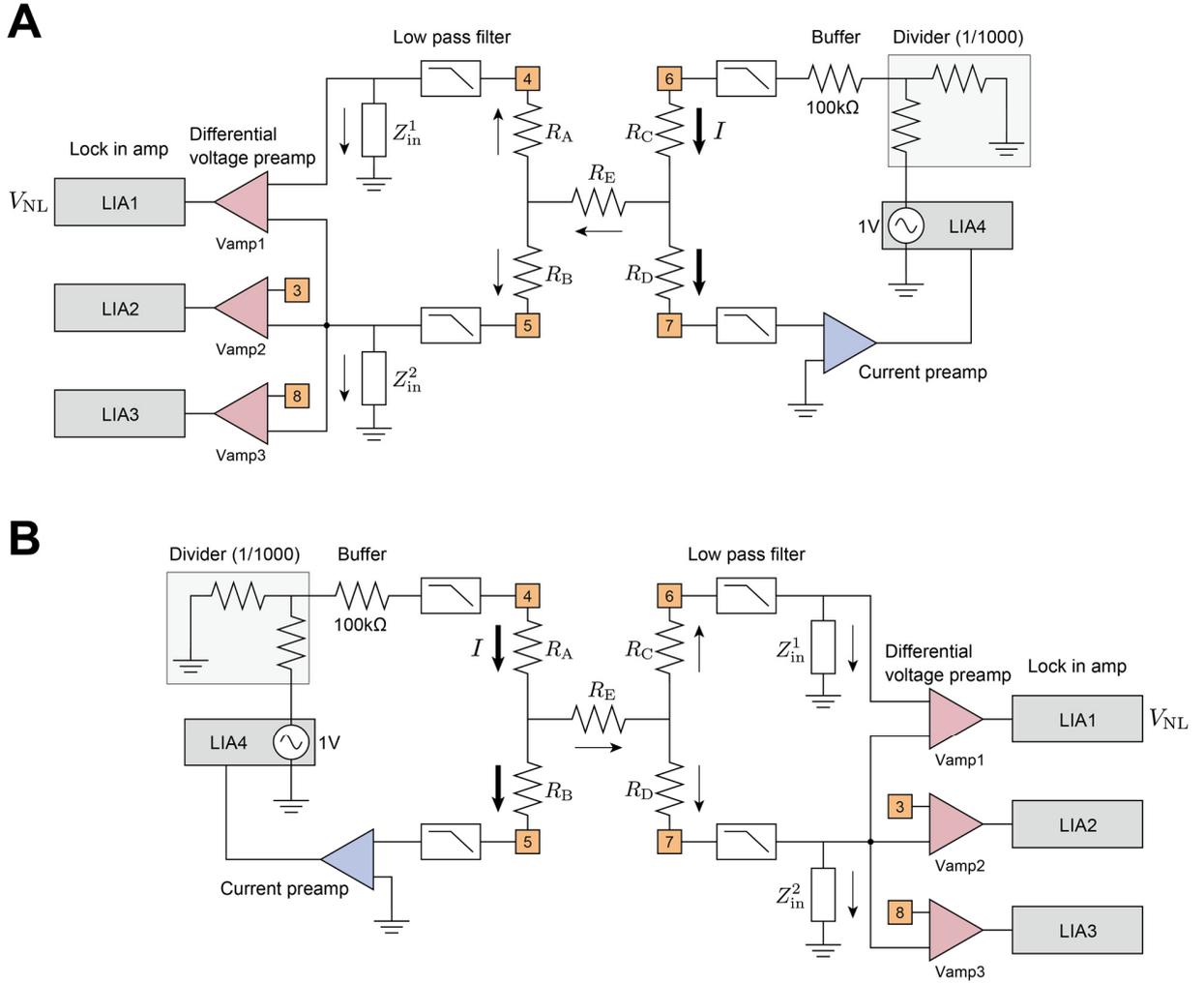

**Fig. S3: Measurement circuit diagram of nonlocal resistance $R_{45,67}$ (A) and $R_{67,45}$ (B) using an AC technique**

To discuss the current leakage effect, total series resistance of the device and the measurement setup is modeled with five resistors of $R_A$ to $R_E$. Each terminal is connected to a commercial low pass filter (BLP-1.9+ from Mini-Circuits) to remove high frequency noise (but not shown for terminal 3 and 8). In (**A**), the AC voltage reduced by a 1/1000 voltage divider is applied with a lock-in amplifier LIA4 to inject an AC current to terminal 6 and 7. A buffer resistor is inserted to suppress possible influences from self-heating in the doped regime (*37*). The injected current is amplified by a current preamplifier and measured with LIA4, while the nonlocal voltage between terminal 4 and 5 is detected, using differential voltage amplifier of Vamp1. The voltage between terminal 3 (8) and 5 is detected using Vamp2 (Vamp3) at the same time. $Z_{in}^1$ and $Z_{in}^2$ denote the effective input impedance including not only the input impedances of the voltage preamplifiers, but also the parasitic capacitance of the measurement system. After amplification of the voltage via Vamp1, 2, and 3, the voltage difference is measured by the lock-in amplifier LIA1, 2, and 3, respectively. The nonlocal voltage signal is then obtained via LIA1. Both in phase (resistive) and out of phase (capacitive) parts of the signal are monitored here. In the measurement of $R_{67,45}$ in (**B**) the electrical connections to the respective terminals are swapped for those in the measurement of $R_{45,67}$.

The finite input impedance effect described above can be present in the nonlocal resistance measurement in the main text as well. Here the impedance $Z_{in}^i$, which causes the leakage current, includes not only the input impedances of the voltage preamplifiers but also parasitic capacitances of the measurement system. For the measurement of $R_{45,67}$, $Z_{in}^i$ are $Z_{in}^1 = 1/(1/R_{in} + i\omega C_p)$ and $Z_{in}^2 = 1/(3/R_{in} + i\omega C_p)$, where $R_{in} = 100\text{M}\Omega$ is the input impedance of the voltage preamplifier and $C_p (= 7.2\text{nF})$ is the parasitic capacitance of the measurement system. Note the dominant source of $C_p$ is the commercial low pass filter. The nonlocal impedance $Z_{45,67}^{leak}$ due to the current leakage to the ground is given by

$$Z_{45,67}^{leak} = \left( \frac{Z_{in}^1}{Z_{in}^1 + R_A} - \frac{Z_{in}^2}{Z_{in}^2 + R_B} \right) R_D$$

$$\sim \left( \frac{R_B}{Z_{in}^2} - \frac{R_A}{Z_{in}^1} \right) R_D$$

$$= \frac{3R_B - R_A}{R_{in}} R_D + i\omega C_p (R_B - R_A) R_D. \qquad (S2)$$

What we measure is the nonlocal impedance $Z_{45,67}^{meas}$ defined as the sum of the nonlocal resistance induced by the valley current and the above described $Z_{45,67}^{\text{leak}}$. The imaginary part of the measured nonlocal impedance is not affected by valley current mediated nonlocal resistance, but only comes from the current leakage effect due to the parasitic capacitance. So the following equation holds:

$$\operatorname{Im} Z_{45,67}^{\text{meas}} = \operatorname{Im} Z_{45,67}^{\text{leak}} = \omega C_{\text{p}} (R_{\text{B}} - R_{\text{A}}) R_{\text{D}}. \quad (S3)$$

We also measure the two-terminal resistance between the terminals 6 and 7,

$$R_{6,7} = R_{\text{C}} + R_{\text{D}}. \quad (S4)$$

Similarly, for the measurement of $R_{67,45}$,

$$Z_{67,45}^{\text{leak}} \sim \frac{3R_{\text{D}} - R_{\text{C}}}{R_{\text{in}}} R_{\text{B}} + i\omega C_{\text{p}} (R_{\text{D}} - R_{\text{C}}) R_{\text{B}}, \quad (S5)$$

and

$$\operatorname{Im} Z_{67,45}^{\text{meas}} = \operatorname{Im} Z_{67,45}^{\text{leak}} = \omega C_{\text{p}} (R_{\text{D}} - R_{\text{C}}) R_{\text{B}}. \quad (S6)$$

We also measure the two-terminal resistance,

$$R_{4,5} = R_{\text{A}} + R_{\text{B}}. \quad (S7)$$

By solving the equations of (S3), (S4), (S6), and (S7) we obtain the following relations for $R_{\text{A}}$, $R_{\text{B}}$, $R_{\text{C}}$, and $R_{\text{D}}$:

$$R_A = \frac{b}{4}\left(3 - \frac{2(c-d)}{\omega C_p ab} - \sqrt{\left(1 + \frac{2(c-d)}{\omega C_p ab}\right)^2 + \frac{8d}{\omega C_p ab}}\right),$$

$$R_B = \frac{b}{4}\left(1 + \frac{2(c-d)}{\omega C_p ab} + \sqrt{\left(1 + \frac{2(c-d)}{\omega C_p ab}\right)^2 + \frac{8d}{\omega C_p ab}}\right),$$

$$R_C = \frac{a}{4}\left(3 + \frac{2(c-d)}{\omega C_p ab} - \sqrt{\left(1 - \frac{2(c-d)}{\omega C_p ab}\right)^2 + \frac{8c}{\omega C_p ab}}\right),$$

$$\text{and } R_D = \frac{a}{4}\left(1 - \frac{2(c-d)}{\omega C_p ab} + \sqrt{\left(1 - \frac{2(c-d)}{\omega C_p ab}\right)^2 + \frac{8c}{\omega C_p ab}}\right). \quad (S8)$$

where $a$, $b$, $c$, and $d$ denotes $R_{6,7}$, $R_{4,5}$, $\text{Im} Z_{45,67}^{meas}$, and $\text{Im} Z_{67,45}^{meas}$, respectively. Therefore the values of $a$, $b$, $c$, and $d$ are all experimentally derived. By using them, we calculated the real part of the nonlocal impedance due to the current leakage effect using the following relations,

$$\text{Re} Z_{45,67}^{leak} = \frac{3R_B - R_A}{R_{in}} R_D, \quad (S9)$$

$$\text{and} \quad \text{Re} Z_{67,45}^{leak} = \frac{3R_D - R_C}{R_{in}} R_B. \quad (S10)$$

By subtracting this real part from the real part of the measured nonlocal impedance, we obtained the nonlocal resistance due to the valley Hall effect as the corrected values of

$$R_{45,67}^{corrected} = \text{Re} Z_{45,67}^{meas} - \text{Re} Z_{45,67}^{leak}, \quad (S11)$$

$$\text{and} \quad R_{67,45}^{corrected} = \text{Re} Z_{67,45}^{meas} - \text{Re} Z_{67,45}^{leak}. \quad (S12)$$

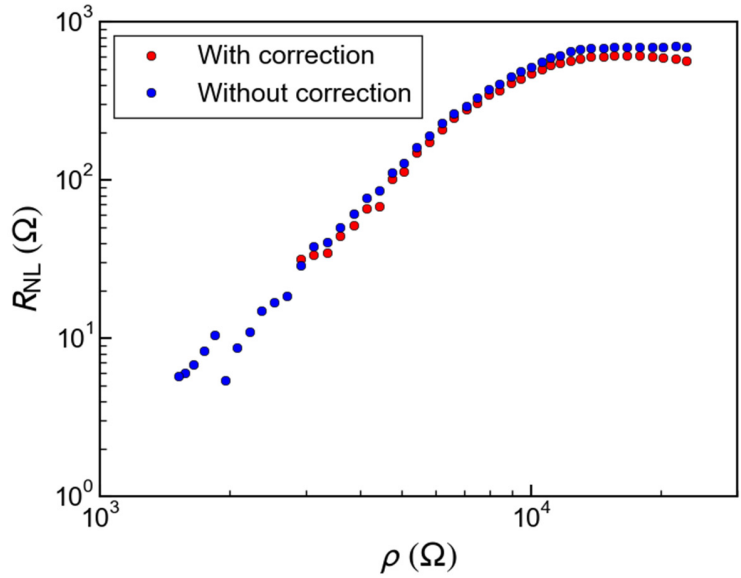

**Fig. S4: Scaling relation of $R_{NL}$ ($= R_{45,67}$) and $\rho$ with and without the current leakage effect correction.**

The red (blue) points show the scaling relation between $R_{NL}$ and $\rho$ with (without) the current leakage effect correction which was obtained for the charge neutrality point by modulating the displacement field at 70K. The blue points are the same as the measurement points in Fig. 4 of the main text.

The correction is usually very small as shown in Fig. S4 for $R_{NL}$ ($= R_{45,67}$) as a typical example. Note, several data points with correction are missing for a small displacement field due to the measurement resolution. In this case the value inside the square root of Eq. S8 becomes negative. After all we see that the current leakage effect does not alter our conclusion. Therefore in the main text, we only showed the data without correction or real part of the measured nonlocal impedance.

## 2.2 Nonlocal resistance measured by swapping injection and detection terminals

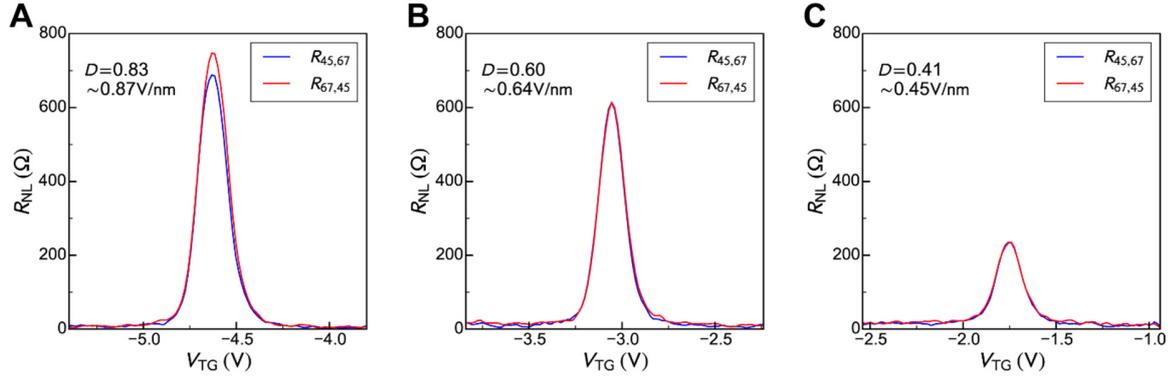

**Fig. S5: Nonlocal resistance measured by swapping injection and detection terminals**

The blue (red) curve was obtained by injecting current from terminal 6 to 7 (4 to 5) and detecting voltage between terminal 4 and 5 (6 and 7) (See Fig 1C.) The top and back gates were biased to modify the carrier density along the green axis of Fig. 2B in the main text. The measurement temperature was fixed at 70K.

Fig. S5 shows the $R_{NL}$ along the green axis of Fig. 2B at 70 K obtained for, $R_{NL} = R_{45,67}$ and $R_{67,45}$ before, and after the injection and detection terminals were swapped, respectively. For almost any gate voltages, the two nonlocal resistances coincide, i.e. $R_{45,67} = R_{67,45}$ as predicted from Onsager reciprocal relations. Slight deviation is probably due to the charge puddle reformation by gate sweeping.

## 2.3 Channel length dependence of nonlocal resistance

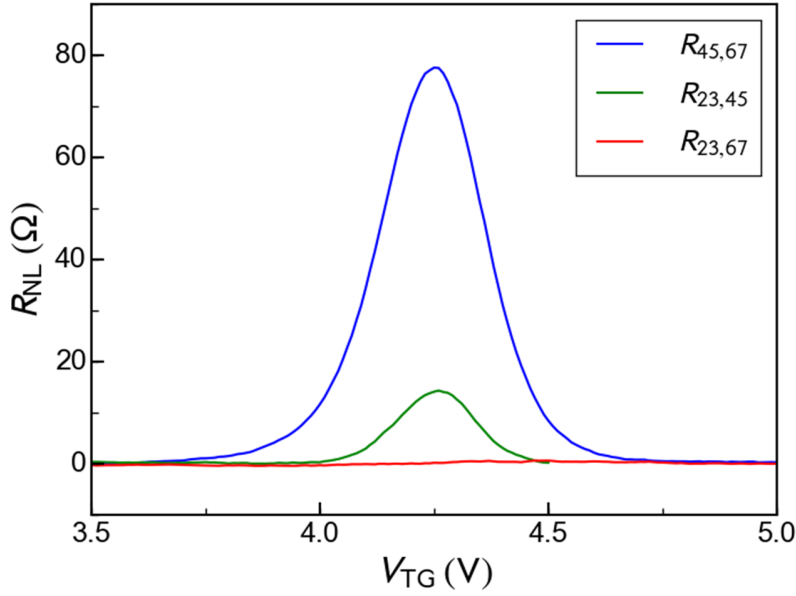

**Fig. S6: Channel length dependence of nonlocal resistance.**

The DC measurement of nonlocal resistance was performed at 73 K for the same BLG device but using different channel length ($L$) Hall bars (see Fig. 1C in the main text). While sweeping $V_{TG}$, $V_{BG}$ was fixed at -50V. Blue, green and red curves are the nonlocal resistances $R_{45,67}$, $R_{23,45}$ and $R_{23,67}$ obtained for $L = 3.5\,\mu m$, $4.5\,\mu m$, and $8\,\mu m$, respectively. For $L = 8\,\mu m$, the signal intensity was smaller than the measurement resolution.

Fig. S6 shows the channel length ($L$) dependence of nonlocal resistance. With increasing the distance or $L$ between the injection and detection terminals, the nonlocal resistance decreases. For $L = 8\,\mu m$, no clear nonlocal resistance signal was observed in our measurement resolution. The reduced nonlocal resistance with increasing $L$ is most likely explained by inter-valley scattering. The dominant source of the inter-valley scattering in graphene has been considered to be edge scattering (*12*, *31*). In this case the scattering length is in the order of the channel width

$W\ (=1\,\mu m)$. Then the observed nonlocal resistance peak may be more significantly reduced for the larger $L$.

**2.4 Derivation of activation energy of local resistivity**

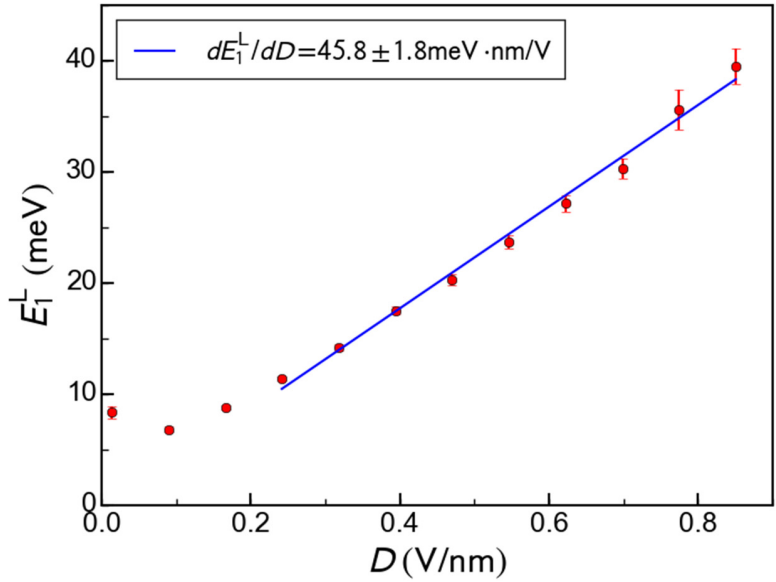

**Fig. S7: Displacement field dependence of local resistivity activation energy $E_1^L$.**

The activation energy $E_1^L$ of the local resistivity is derived from the fitting of the temperature dependence presented in Fig. 3A of the main text. The blue line is a linear fit for the data points.

The temperature dependence of the maximum local resistivity for each fixed displacement field $D$ was fit to the double exponential function of Eq. 4 in the main text. When the data points are plotted with respect to the inverse temperature, the intervals of the data points are not uniform as shown in Fig. 3A. To extract the activation energy $E_1^L$, the data points were not weighted according to the density of the data points. This is because the data points in the high temperature regime are dense and the fitting without the weighting allows more correct estimation of $E_1^L$. The extracted $E_1^L$ for each $D$ is shown in Fig. S7. This $E_1^L$ corresponds to $\Delta$ (half of the band gap) and linearly increases with $D$, indicating the band gap size increased up to $2\Delta = 80\text{meV}$ for the large $D$. On the other hand, $E_1^L$ saturates for the region of small $D$. This saturation is attributed to the mobility edge effect (*17*).

.Note that to extract the crossover temperature in Fig. 3A, the data points were weighted according to the density of the data points.

## 2.5 Derivation of activation energy of nonlocal resistance and crossover behavior of nonlocal resistance in displacement field dependence

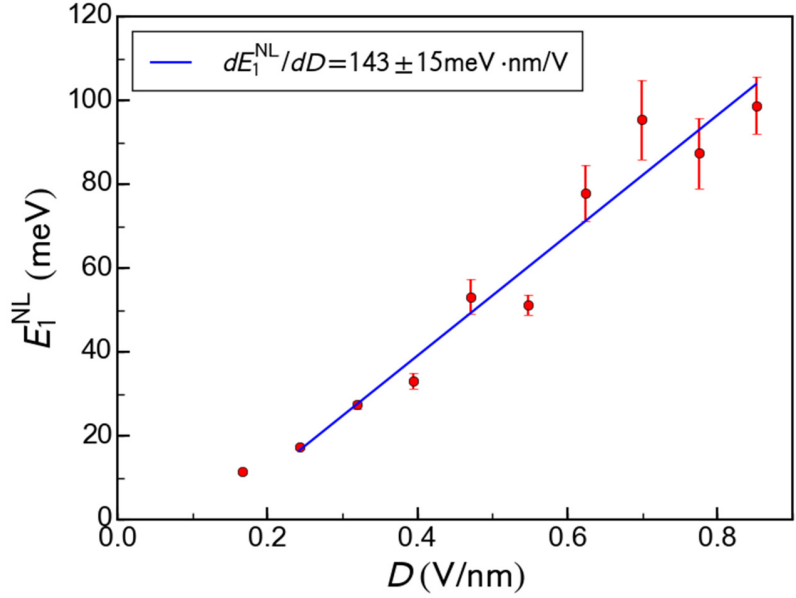

**Fig. S8: Displacement field dependence of the nonlocal resistance activation energy $E_1^{NL}$.**

The activation energy $E_1^{NL}$ of the nonlocal resistance is derived from the fitting of the temperature dependence presented in Fig. 3B of the main text. The blue line is the linear fit for the data points.

The temperature dependence of the maximum nonlocal resistance for each fixed $D$ was fit to the double exponential function of Eq. 5 in the main text. In the same way as for the local resistivity, the data points were weighted to extract the crossover temperature but not weighted to extract $E_1^{NL}$. Fig. S8 shows the extracted activation energy $E_1^{NL}$ for each $D$. $E_1^{NL}$ scales linearly with $D$. However, the extrapolation of the fitting line in blue to $D=0$ shows a small offset energy.

The origin of this offset is not yet understood. By applying Eq 4 and Eq 5 to the cubic scaling in Eq 6: $R_{NL} \propto \rho^3$, we expect the relationship between activation energies to be: $E_1^{NL} = 3E_1^{L}$. In fact, the ratio of the slopes of the fitting lines is 3.13 ± 0.36 for $D \gtrsim 0.2\text{V/nm}$ meaning: $\frac{dE_1^{NL}}{dD} \sim 3\frac{dE_1^{L}}{dD}$. However, due to the zero offset of $E_1^{NL}$, the absolute relation $E_1^{NL} = 3E_1^{L}$ does not hold for the small $D$. This observation also supports that the observed nonlocal resistance comes from the valley current mediated nonlocal transport.

Note that in (12), similar activation energy between local and nonlocal resistance was found, while our results show discrepancy between them.

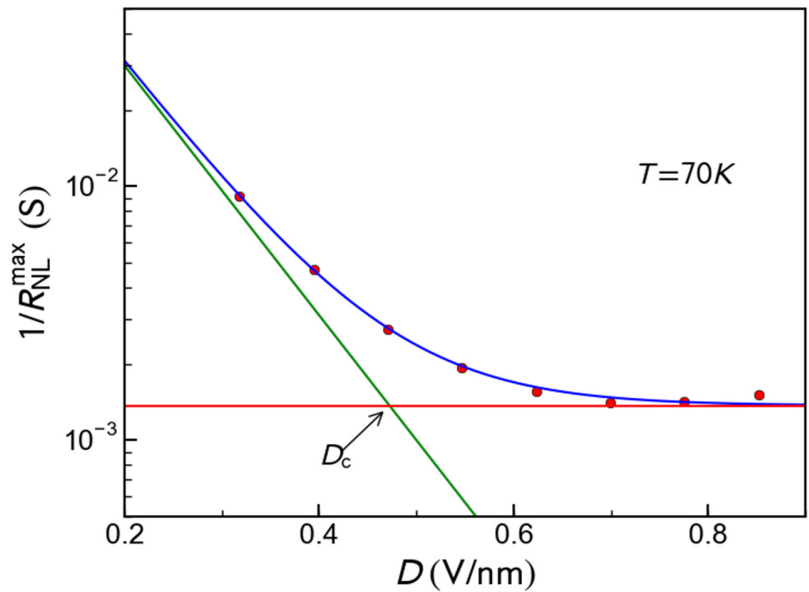

**Fig. S9: Fitting example of the displacement field dependence of nonlocal resistance.**

The inverse of the maximum nonlocal resistance $1/R_{NL}^{max}$ is obtained at each $D$ when the carrier density is varied at a fixed temperature.

The crossover behavior we observed for the $T$ dependence of $R_{\text{NL}}$ in Fig. 3B of the main text was also observed for the $D$ dependence of $R_{\text{NL}}$. Fig. S9 shows the $D$ dependence of $1/R_{\text{NL}}^{\max}$ obtained at 70K. The data points were well fitted to the function:

$$\frac{1}{R_{\text{NL}}^{\max}} = \frac{1}{R_1}\exp\left(-\frac{D}{D_1}\right) + \frac{1}{R_2}. \quad (S13)$$

In this figure we define the crossover displacement field $D_c$ at the point where the first term (green line) and the second term (red line) of Eq. S13 intersect.

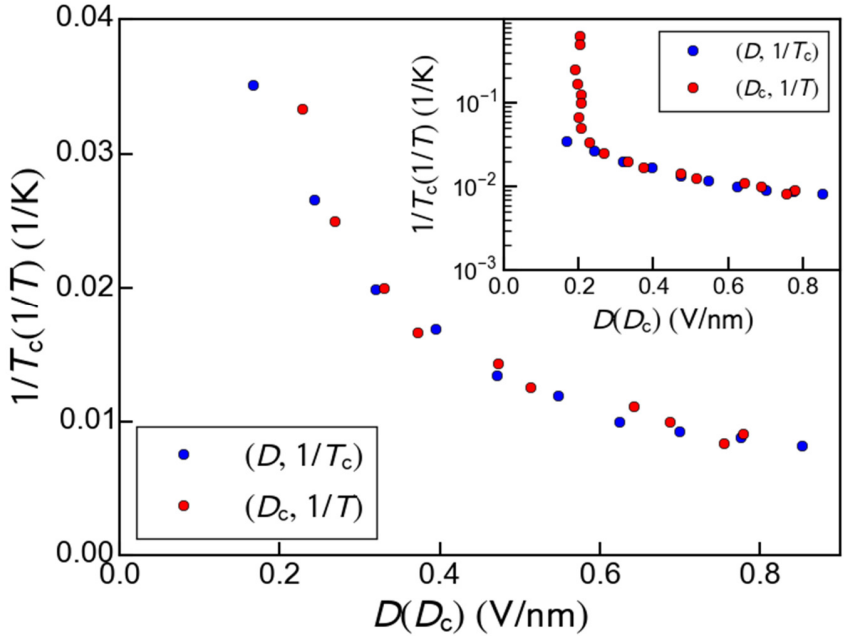

**Fig. S10: Relation between nonlocal resistance crossover temperature and crossover displacement field.**

The blue points show the $1/T_c$ vs. $D$ where $T_c$ was obtained from the $T$ dependence of $1/R_{\text{NL}}^{\max}$ for each fixed $D$. The red points show the $D_c$ vs. $1/T$ where $D_c$ was obtained from the $D$ dependence of $1/R_{\text{NL}}^{\max}$ for each fixed $T$. Some red points where $D$ is around 0.2V/nm are exceeding the range. (Inset) All data points are shown in the log scale. Here the deviation of

$D_c - 1/T$ curve from the $D - 1/T_c$ curve was observed where $D$ is around 0.2V/nm.

$D_c$ extracted for each $T$ in this way is plotted in Fig. S10 together with $T_c$ extracted for each $D$ (see Fig. 3C in the main text). The $D - 1/T_c$ and $D_c - 1/T$ curves show good correspondence. We assume that there is a crossover phase boundary of nonlocal resistance in the $D - 1/T$ plane, necessitating the existence of critical points $D_c(T)$ and $T_c(D)$. So good correspondence between the $D - 1/T_c$ curve and the $D_c - 1/T$ curve, as obtained in Fig. S10, supports the validity of our estimation of $T_c$ in the main text. However, for a smaller displacement field around 0.2V/nm, the deviation of the $D_c - 1/T$ curve from the $D - 1/T_c$ curve was observed (see Fig. S10 inset). The reason for this deviation is yet to be revealed.

..

## 2.6 $R_{NL} - \rho$ scaling relation obtained near the charge neutrality point

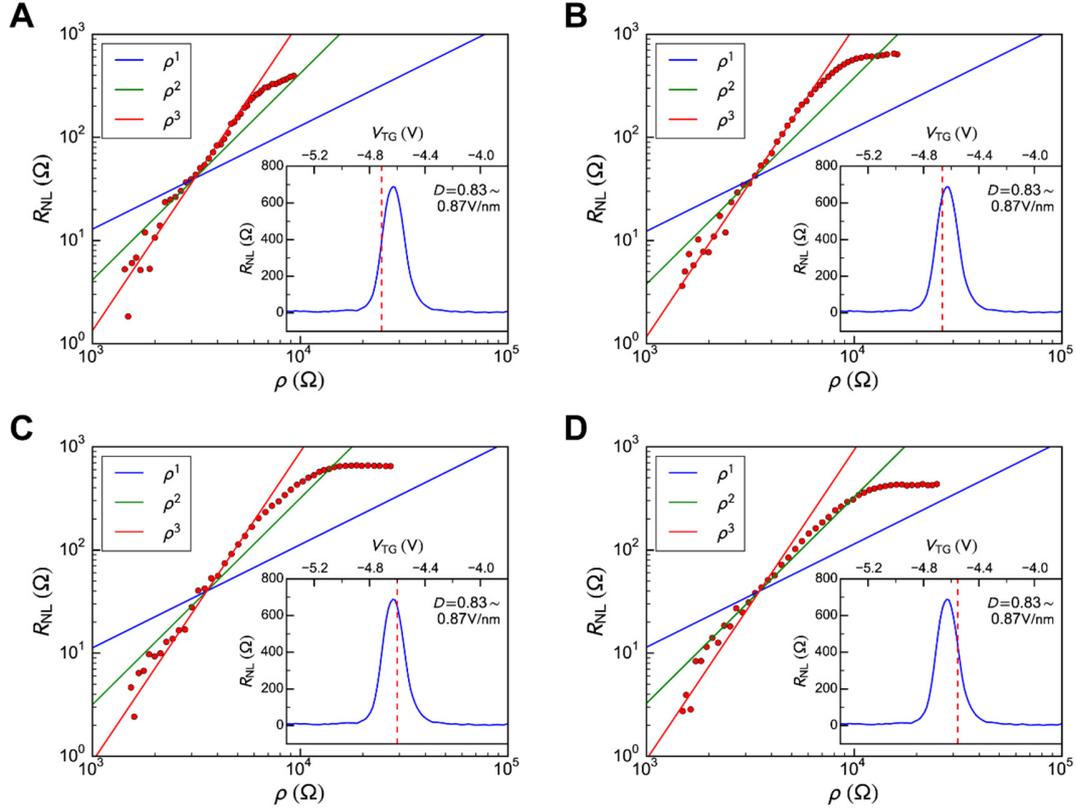

**Fig. S11: Scaling relation of $R_{NL}$ vs. $\rho$ obtained near the charge neutrality point.**

Each data point of $R_{NL}$ vs. $\rho$ was obtained from Fig. 2A and B in the main text by isolating only the $D$ dependence of $R_L$ and $R_{NL}$ (taking a cut along the red axis) for four different gate configurations (**A** to **D**), all near the charge neutrality point. (Insets) The blue curve shows $R_{NL}$ data cut along the green axis in Fig. 2B. Only the case for the highest $D$ is shown. The $R_{NL}$ vs. $\rho$ in each main panel was obtained such that for every $D$, the resistance data was taken at the red broken line.

In the main text, a clear scaling relation between $R_{NL}$ and $\rho$ at the charge neutrality point and $T = 70K$ is shown in Fig. 4. A similar scaling relation is obtained for several gate configurations near (but away from) the charge neutrality point as shown in Fig. S11. Fig. S11A, B, C show cubic scaling in the smaller displacement field regime.

In Fig. S11D, the scaling starts to deviate from the cubic scaling. An extrinsic mechanism of the valley Hall effect that gives a deviation from what is expected from the intrinsic mechanism model is possible (*10*).

## 2.7 Derivation of nonlocal resistance scaling for the edge dominant transport

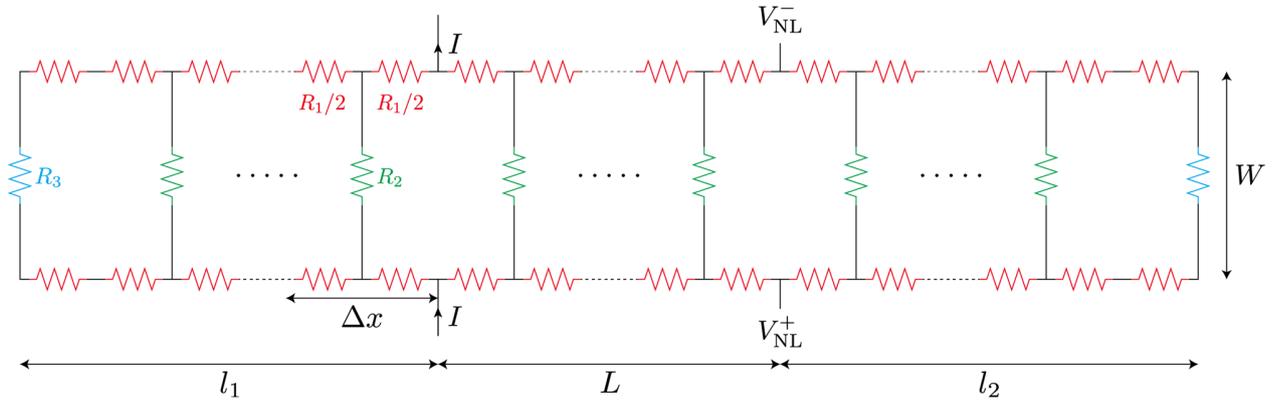

**Fig. S12: Resistor network circuit model for nonlocal edge transport.**

Uniform edge resistivity $\rho_{\text{edge}}$, and bulk resistivity $\rho_{\text{bulk}}$ are assumed to model the edge transport. Current $I$ is injected from the lower side of the network circuit and ejected from the upper side. The nonlocal voltage $V_{\text{NL}} = V_{\text{NL}}^{+} - V_{\text{NL}}^{-}$ is measured between two points each separated by $L$ from the current injection and ejection terminal, respectively. The distance from the current injection (voltage probe) terminals to the left (right) boundary of the network circuit is $l_1$ ($l_2$) and the circuit width from the current injection to ejection is $W$. The circuit is divided into sections of length $\Delta x$, and we finally take the limit of $\Delta x \to 0$. The resistor at each section is $R_1 / 2 = \rho_{\text{edge}} \Delta x / 2$ in red, $R_2 = \rho_{\text{bulk}} W / \Delta x$ in green. In the experimental system, the left and right boundaries are highly doped with the back gate voltage. So the boundary resistance $R_3$ in cyan is much smaller than the bulk and the edge resistance, and therefore approximated as $R_3 = 0$.

To take the nonlocal transport mediated by localized edge states into account, we employed a resistor network model. This kind of resistor network model was previously used to describe the nonlocal transport in an imperfect 2D topological insulator which has a dissipative edge channel and the residual bulk conduction (*38*). Here we assume the case where the edge transport is dominant and the lateral transport via the bulk states is negligible because we already discussed the opposite case where the nonlocal resistance is given by van der Pauw formula for the bulk dominant transport in the main text. Fig. S12 shows the resistor network circuit modeled here. By using transmission matrices and taking the limit of $\Delta x \to 0$ (see Fig. S12), we obtain

$$R_{NL} = \frac{\sinh(l_1/\lambda)\sinh(l_2/\lambda)}{\sinh((L+l_1+l_2)/\lambda)}\sqrt{2\rho_{bulk}\rho_{edge}W}, \quad (S14)$$

with $\lambda = \sqrt{\rho_{bulk}W/2\rho_{edge}}$.

For $\lambda \ll l_1, l_2$, Eq. S14 is approximated by

$$R_{NL} = \sqrt{\frac{\rho_{bulk}\rho_{edge}W}{2}}\exp\left(-\frac{L}{\lambda}\right), \quad (S15)$$

while for $\lambda \gg l_1, l_2$, it is approximated by

$$R_{NL} = \frac{2l_1 l_2}{L+l_1+l_2}\rho_{edge}. \quad (S16)$$

On the other hand, the local resistance is given by the parallel connection of the upper and lower edge resistances as,

$$R_L = \frac{\rho_{edge}L}{2}. \quad (S17)$$

In the analysis in the main text, we defined two dimensional resistivity as $\rho = R_L/(L/W)$, and

then it is given by

$$\rho = \frac{1}{2}\rho_{edge}W. \quad (S18)$$

So for $\lambda \ll l_1, l_2$, from Eqs. S15 and Eq. S18, the relation between $R_{NL}$ and $\rho$ is given by the following formula.

$$R_{NL} = \sqrt{\rho_{bulk}\rho}\exp\left(-\frac{2L}{W}\sqrt{\frac{\rho}{\rho_{bulk}}}\right). \quad (S19)$$

In Eq. S19 $\rho_{bulk}$ also depends on $D$, so that the actual scaling relation of $R_{NL}$ vs. $\rho$ should be more complicated.

On the other hand, when $\lambda \gg l_1, l_2$, from Eqs. S16 and Eq. S18, the scaling relation between $R_{NL}$ and $\rho$ is given by

$$R_{NL} = \frac{4l_1 l_2}{(L+l_1+l_2)W}\rho \propto \rho. \quad (S20)$$

In either case, the scaling relation obtained here cannot be applied for the observed cubic scaling. Therefore we exclude the possibility of the edge mediated nonlocal transport in the present study.

## 2.8 Self-consistent model of valley current mediated nonlocal transport

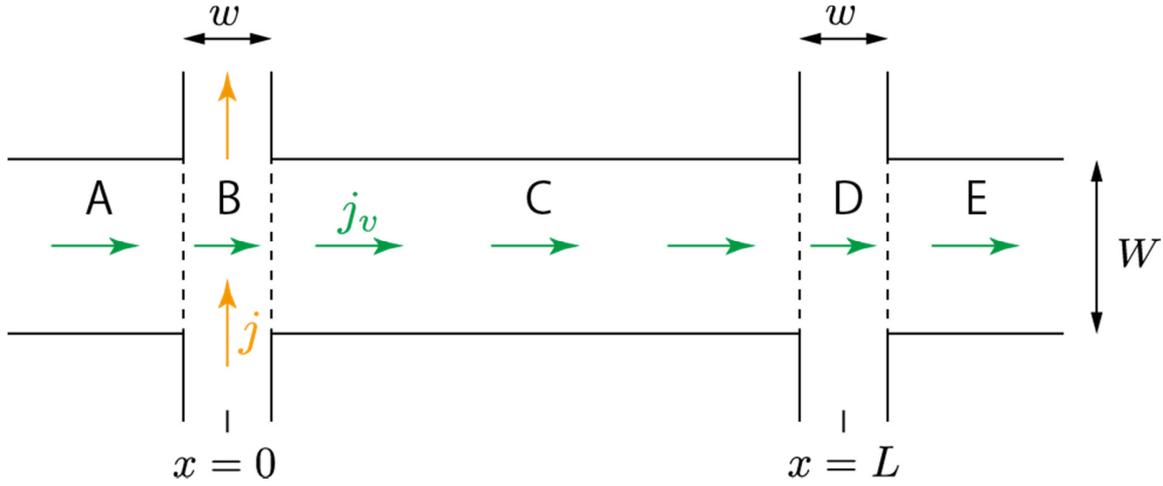

**Fig. S13: Schematic picture of the nonlocal transport mediated by valley current.**

Here we assume a Hall bar which is infinitely-long in $x$ direction. Two probes separated by $L$ are attached to both sides of the central channel whose width is $W$. The probe width is $w$. The Hall bar is separated into five regions, A, B, C, D, and E. Charge current is injected into the region B and nonlocal voltage is measured in the region D.

The nonlocal resistance which results from the combination of the spin Hall effect and the inverse spin Hall effect was previously calculated for a sequential conversion picture (*23*). In this picture, an injected current induces a transverse spin current and the spin current induces a nonlocal voltage. This is correct when the spin Hall angle $\alpha_{\mathrm{SH}} = \sigma_{xy}^{\mathrm{SH}} / \sigma_{xx}$ fulfills the condition of $\alpha_{\mathrm{SH}} \ll 1$. In (*12*), this picture was applied to the valley current mediated nonlocal transport and the nonlocal resistance observed for a doped region where $\alpha_{\mathrm{VH}} = \sigma_{xy}^{\mathrm{VH}} / \sigma_{xx} \leq 1$ holds was analyzed. In contrast, we discuss the nonlocal transport in the gap. If we only consider the intrinsic valley Hall effect, the valley Hall conductivity in the gap is $\sigma_{xy}^{\mathrm{VH}} = 4e^2 / h$ as discussed in the main text. So for a large displacement field, the valley Hall angle exceeds one. In such a case, the sequential conversion picture fails. We need to treat the problem in a self-consistent way with a certain boundary condition.

Fig. S13 shows the schematic picture of the nonlocal transport mediated by the valley current which holds for our system. Chemical potential difference $\delta\mu_v$ between the chemical potentials, $\mu_K$ and $\mu_{K'}$ of two valleys K and K′, which can be interpreted as the valley voltage, is defined as $\delta\mu_v = \mu_K - \mu_{K'}$. It follows that the diffusion equation is derived as

$$\frac{\partial^2}{\partial x^2}\delta\mu_v^i = \frac{1}{(l_v^i)^2}\delta\mu_v^i, \quad (S21)$$

where the suffix $i$ corresponds to the region index and $l_v^i$ is the inter-valley scattering length. The inter-valley scattering should mainly be caused by the edge scattering, so we assume that $l_v^A = l_v^C = l_v^E = l_v$ and $l_v^B = l_v^D = \infty$. The relation between the charge current density $j_c$ in the $y$ direction, the valley current density $j_v$ in the $x$ direction, the electric field $E$ in the $y$ direction and the gradient $\frac{\partial}{\partial x}\delta\mu_v$ of the valley voltage in the $x$. direction is given by the conductance matrix:

$$\begin{pmatrix} j_c^i \\ j_v^i \end{pmatrix} = \begin{pmatrix} \sigma_{xx} & -\sigma_{xy}^{VH} \\ \sigma_{xy}^{VH} & \sigma_{xx} \end{pmatrix} \begin{pmatrix} E^i \\ \frac{1}{2e}\frac{\partial}{\partial x}\delta\mu_v^i \end{pmatrix}, \quad (S22)$$

where we assume uniform $\sigma_{xx}$ and $\sigma_{xy}^{VH}$. We neglect the current dispersion effect, so the charge current density is zero except for the region B. In the region B, we assume the uniform charge current density $j_c^B = j$. At each interface of neighboring regions in Fig. S13, the valley voltage and the valley current density are connected continuously. We assume the valley voltage is zero at the infinitely distant points, namely $\delta\mu_v^A(x=-\infty)=0$ and $\delta\mu_v^E(x=\infty)=0$. By solving Eqs. S21 and S22 self-consistently with these boundary conditions, we obtain the following nonlocal resistance formula:

$$R_{NL} = \frac{W}{2l_v} \frac{\left(\sigma_{xy}^{VH}\right)^2}{\sigma_{xx}\left(\sigma_{xx}^2+\left(\sigma_{xy}^{VH}\right)^2\right)} \exp\left(-\frac{L-w}{l_v}\right) \frac{1}{\left(1+\frac{w}{2l_v}\right)^2 - \left(\frac{w}{2l_v}\right)^2 \exp\left(-2\frac{L-w}{l_v}\right)}. \quad (S23)$$

So for $\alpha_{VH} \ll 1$, Eq. S23 is approximated by

$$R_{NL} = \frac{W}{2l_v} \frac{\left(\sigma_{xy}^{VH}\right)^2}{\sigma_{xx}^3} \exp\left(-\frac{L-w}{l_v}\right) \frac{1}{\left(1+\frac{w}{2l_v}\right)^2 - \left(\frac{w}{2l_v}\right)^2 \exp\left(-2\frac{L-w}{l_v}\right)}, \quad (S24)$$

and the scaling relation between $R_{NL}$ and $\rho = 1/\sigma_{xx}$ is $R_{NL} \propto \rho^3$.

Further taking the limit of $w \to 0$, Eq. S24 becomes

$$R_{NL} = \frac{W}{2l_v} \frac{\left(\sigma_{xy}^{VH}\right)^2}{\sigma_{xx}^3} \exp\left(-\frac{L}{l_v}\right). \quad (S25)$$

This reproduces the nonlocal resistance formula calculated in the sequential conversion picture in (*23*).

On the other hand, when $\alpha_{VH} \gg 1$, Eq. S23 is approximated by the following formula.

$$R_{NL} = \frac{W}{2l_v} \frac{1}{\sigma_{xx}} \exp\left(-\frac{L-w}{l_v}\right) \frac{1}{\left(1+\frac{w}{2l_v}\right)^2 - \left(\frac{w}{2l_v}\right)^2 \exp\left(-2\frac{L-w}{l_v}\right)}. \quad (S26)$$

So the scaling relation between $R_{NL}$ and $\rho$ is $R_{NL} \propto \rho$. But this does not explain the saturation of the nonlocal resistance obtained for a large displacement field (see Fig. 4 in the main text).